\documentclass[twocolumn,aps,showpacs,preprintnumbers,nofootinbib,
superscriiptaddress]{revtex4}
\usepackage{graphicx}
\usepackage{dcolumn}
\usepackage{bm}
\begin{document}

\title{ENHANCED OPTICAL COOLING OF PARTICLE BEAMS IN STORAGE RINGS}

\author{E.G.Bessonov, Lebedev Phys. Inst. RAS, Moscow, Russia \\
 A.A.Mikhailichenko, Cornell University, Ithaca, NY, U.S.A.}

                       \begin{abstract}
The problem of enhanced optical cooling (EOC) of particle beams in 
storage rings beyond the Robinson's damping criterion is discussed.  
                       \end{abstract}


\maketitle

            \section{Introduction}
For any particle in a storage ring, the change of the square of its 
amplitude of betatron oscillations caused by sudden energy change 
$\delta E$ in smooth approximation is determined by the equation

\vskip -5mm
\begin{equation}
\label{eq1}
\delta A_x^2 = - 2x_{\beta ,0} \delta x_\eta + (\delta x_\eta )^2,
\end{equation}
\vskip -1mm
\noindent
where $x_{\beta ,0} $ is the initial particle deviation from it's closed
orbit; $\delta x_\eta = \eta _x \beta ^{ - 2}(\delta E / E)$ --is the 
change of it's closed orbit position; $\eta _x $ is the dispersion 
function in the storage ring; $\beta $ is the normalized velocity. In 
the approximation $\vert \delta x_\eta \vert < 2\vert x_{\beta 0} \vert 
< 2A$ the amplitude will be decreased, if the product $x_{\beta ,0} 
\delta x_\eta > 0$. Based on this observation, two schemes of EOC of 
particle beams based on external selectivity were suggested in [1] for 
unbunched beam (RF turned off).  Pick-up and kicker undulators, laser 
and optical systems in these schemes of cooling are similar to 
corresponding systems in the method of the optical stochastic cooling 
(OSC) [2]. The possibility to enhance OSC by screening the radiation 
from some part of the beam was mentioned also in [3].

In the first scheme of EOC two or more identical undulators are installed in
different straight sections of a storage ring at a distance determined by a
betatron phase advance for the lattice segment $(2p + 1)\pi $ between
pick-up and first kicker undulator and $2{p}'\pi $ between next kicker
undulators; where $p,\;{p}'$ = 1,2,3... are integer numbers. Undulator
Radiation Wavelets (URW) are emitted by a particle in the pick-up undulator
passed through an optical system with movable screens located on the image
plane of the particle's beam. Then this radiation is amplified and passes
through the following kicker undulators together with the particle.

First, the screens in the optical system open the way for URW emitted in the
pick-up undulator by particles with higher energies and higher positive
deviations $x_\beta > 0$ from their closed orbits. The beam of amplified URW
in the kicker undulators, in this case is similar to the moving prototype of
the target $T_2$ considered in [1], [4] if definite phase conditions 
are fulfilled in the optical system to inject particles in the kicker 
undulators at decelerating phases. If the betatron phase advance for 
the lattice segment between pick-up and kicker undulators is $(2p + 
1)\pi $ and the deviation of the particle in the pick-up undulator 
$x_\beta > 0$, then the deviation of the particle in the kicker 
undulators comes to be $x_\beta < 0$. In this case the energy loosed by 
particles is accompanied by a decrease in both energy spread and 
amplitudes of betatron oscillations of the beam.  So the EOC is going 
both in the longitudinal and transverse degrees of freedom. After the 
screen will open images of all particles of the beam the optical system 
must be switched off. Then the cooling process can be repeated.

Second modification of EOC method uses the pick-up undulator followed by
even number of kicker undulators installed in straight sections of a storage
ring. The distances between neighboring undulators are determined by the
phase advance equal to $(2p + 1)\pi $. In this scheme, the deviations of
particles in undulators "$i$" and "$i$+1" are $x_{\beta _i } = - x_{\beta _{i + 1}
} $ and that is why the decrease of energy of particles in undulators does
not lead to change of the particle's betatron amplitude at the exit of the
last undulator. So it leads the cooling of the particle beam in the
longitudinal coordinate only.

A third modification of EOC method can be suggested. It uses the pick-up
undulator and even number of kicker undulators installed in straight
sections of a storage ring at distances determined by a phase advance $(2p +
1)\pi $ between the pick-up and next neighboring kicker undulators as well.
If the particle decreases its energy in odd undulators and increases it in
even ones, then the change of the energy of the particle in undulators leads
to decrease of their betatron amplitudes and do not lead to change of their
energy at the exit of the last undulator. In this case cooling of the
particle beam is going in the transverse coordinate only.

The wavelets of UR emitted by a particle in the pick-up undulator after
amplification in the optical amplifier interact efficiently with the
particle in the kicker undulators. Radiation from one particle does not
disturb trajectories of other particles if an average distance between
particles in a longitudinal direction is more, than the length of the URW,
$M\lambda _{UR} $, where $M$ is the number of the undulator periods; $\lambda
_{UR} $ is the wavelength of the emitted undulator radiation (UR). This 
case is named ``single particle in the sample''. It corresponds to the 
beam current

\vskip -5mm
\begin{equation}
\label{eq2}
i < i_c = \frac{Zec}{M\lambda _{UR} } = \frac{4.8 \cdot 10^{ - 9}Z}{M\lambda
_{UR} }[A]
\end{equation}
\vskip -2mm
If overlapping of other particles with URW occurs (more than one particle in
the sample), then amplified URW does not disturb the energy and amplitudes
of betatron oscillations of other particles of the beam in the first
approximation. It leads to an increase of their amplitudes in the second
approximation because of the stochasticity of the initial phase of the URW
for other particles.

In the second method the URW emitted by a particle in the pick-up undulator
and amplified in the optical amplifier interact efficiently with the same
particle in the kicker undulators and do not disturb amplitudes of betatron
oscillations of other particles independently of the average distance
between particles.

In the third method the amplified URW do not disturb the energy and
amplitudes of betatron oscillations of other particles of the beam in the
first approximation. It leads to a week increase in theirs betatron
amplitudes in the second approximation because of the stochasticity of the
initial phase of the URW for other particles.

In these schemes of cooling at first approximation, the degree of cooling of
high current beams is higher, if the transverse dimensions of the URW in the
kicker undulators are less then the transverse total (dispersion + betatron)
dimensions of the being cooled particle beam as in this case the particles
outside the URW do not interact with the URW and the characteristic current
(\ref{eq2}) is increased in the ratio of the areas of the particle beam and URW. It
means that high dispersion- and beta- functions in the straight section of
the storage ring have to be used at the location of the pick-up and kicker
undulators.

The considered schemes are of great interest for cooling of fully stripped
ion, proton and muon beams. Laser cooling, based on nuclear transitions has
problems with low-lying levels [5]. Enhanced optical cooling of heavy ions,
on the level with optical stochastic cooling, is the most efficient. In this
case the emitted power $\cong Z^2$, where $Z$ is the atomic number [6].

\section{THE RATE OF COOLING}

The total energy radiated by a relativistic particle traversing a given
undulator magnetic field $B$ of finite length is given by

\vskip -5mm
\begin{equation}
\label{eq3}
E_{tot} = \textstyle{2 \over 3}r_p^2 \overline {B^2} \gamma ^2M\lambda _u ,
\end{equation}
\vskip -2mm
\noindent
where $\overline {B^2} $ is an average square of magnetic field along the
undulator length $M\lambda _u $; $\gamma $ is the relativistic factor; 
$r_p = Z^2e^2 / M_p c^2$ is the classical radius of the particle; $Z = 
1$ for electrons, protons and muons. All other symbols have theirs 
standard meanings. For a plane harmonic undulator, $\overline {B^2} = 
B_0^2 / 2$, where $B_{0}$ is the peak of the undulator field. For 
helical undulator $\overline {B^2} = B_0^2 $.

The expression for the wavelength of the undulator radiation is

\vskip -5mm
\begin{equation}
\label{eq4}
\lambda _{UR,k} = \frac{\lambda _u }{2k\gamma ^2}(1 + K^2 + \vartheta ^2),
\end{equation}
\vskip -2mm
\noindent
where $\lambda _{UR,k} $ is the wavelength of the $k^{th}$ harmonic of UR;
$\vartheta = \gamma \theta $; $\theta $, the azimuth circle angle and
$K$ is the deflection parameter given by

\vskip -5mm
\begin{equation}
\label{eq5}
\sqrt {\overline {K^2} } = \frac{Ze\sqrt {\overline {B^2} } \lambda _u
}{2\pi M_p c^2}.
\end{equation}
\vskip -2mm
The number of the equivalent photons in the URW, according to (\ref{eq3})
-- (\ref{eq5}) becomes

\vskip -5mm
\begin{equation}
\label{eq6}
N_{ph} = \frac{E_{tot} }{\hbar \omega _{1,\min } } = \textstyle{2 \over
3}\pi \alpha MZ^2\overline {K^2},
\end{equation}
\vskip -2mm
\noindent
where $\omega _{1,\max } = 2\pi c / \lambda _{UR,1} \vert _{\theta = 0} $.

In the regime of small deflection parameter $K < 1$, the spectrum of
radiation emitted in the undulator having harmonically varying transverse
magnetic field, is given by

\vskip -5mm
\begin{equation}
\label{eq7}
\frac{dE}{d\xi } = E_{tot} f(\xi ),
\end{equation}
\vskip -2mm
\noindent
where $f(\xi ) = 3\xi (1 - 2\xi + 2\xi ^2), \quad \xi = \frac{\lambda
_{UR,1,\min } }{\lambda }$, ($0 \le \xi \le 1)$, $\int {f(} \xi )d\xi = 
1, \quad \lambda _{UR,\min } = \lambda _u (1 + K^2) / 2\gamma ^2$.

The bandwidth of the UR emitted at a given angle $\theta $

\vskip -5mm
\begin{equation}
\label{eq8}
\frac{\Delta \omega }{\omega } = \frac{1}{kM}.
\end{equation}
\vskip -2mm
Below we accept a Gaussian distribution for the URW, its Rayleigh 
length $Z_R = 4\pi \sigma _w^2 / \lambda _{UR,1.\min } = M\lambda _u $; 
where $\sigma _w $ is the rms waist size. In this case $ \sigma _w = 
\sqrt {Z_R \lambda _{UR,1,\min } / 4\pi }$.

The rms electric field strength $E_w $ of the wavelet in the kicker
undulator

\vskip -5mm
\begin{equation}
\label{eq9}
E_w = \sqrt {\frac{2E_{tot} }{\sigma _w^2 M\lambda _{UR,1,\min } }} =
\frac{2\sqrt {2\overline {B^2} } \gamma ^2r_p }{\sqrt 3 \sigma _w }.
\end{equation}
\vskip -2mm
The rate of the energy loss for particles in the amplified URW is

\vskip -5mm
\begin{equation}
\label{eq10}
P_{loss} = eE_w M\lambda _u \beta _{ \bot m} f \cdot N_{kick} \sqrt {\alpha
_{ampl} } ,
\end{equation}
\vskip -2mm
\noindent
where $\beta _ \bot = K / \gamma $ is the maximum deflection angle of a
particle from the direction of its closed orbit; $f$ is the revolution
frequency; $N_{kick} $ is the number of kicker magnets; $\alpha _{ampl} $ is
the gain in optical amplifier.

The damping time for the particle beam in the longitudinal degree of freedom
is

\vskip -5mm
\begin{equation}
\label{eq11}
\tau = \Delta E_b / P_{loss} ,
\end{equation}
\vskip -2mm
\noindent
where $\Delta E_b $ is the energy spread of the particle beam.

According to (\ref{eq11}), the damping time of the particle beam in the longitudinal
plane is shorter, proportionally to its energy spread (not to the initial
energy of particles). Moreover, because of non-exponential decay of its both
energy and angular spreads the degree of cooling is much higher than 1/e
reduction of the beam emittance [1].

If in the pick-up method of cooling the screen of the optical system will
open images of all particles of the beam and at this moment it will be
stopped, the open particles will continue to loose their energies up to the
moment when they will be displayed inward to the distances corresponding to
overlapping their URW by the screen. After this time all particles will stay
at a threshold energy with the energy spread determined by the jump of the
particle energy $\Delta E_{loss} = P_{loss} \sqrt {n_c } / f$, where $n_c =
(i / i_c )(\sigma _{URW} / \sigma _b )$ is the number of particles in a
sample; $\sigma _{URW} $ is the transverse area occupied by the URW and
$\sigma _b $ is the area occupied by the particle beam.

The minimum rms transverse dimension of the beam $\sigma _x $ is determined
by jumps $\delta x_\eta $ of the closed orbit of particles. According to
(\ref{eq1}),

\vskip -5mm
\begin{equation}
\label{eq12}
\overline {A_x^2 } = (\delta x_\eta )^2N_c ,
\end{equation}
\vskip -2mm
\noindent
where $N_c = (\Delta E_b / \Delta E_{loss} )n_c $ is the number of
interactions of particles with URWs. In this case the value

\vskip -5mm
$$
\sigma _x = \sqrt {\overline {A_x^2 } } = \delta x_\eta \sqrt {\frac{\Delta
E_b }{\Delta E_{loss} }n_c } =
$$
\begin{equation}
\label{eq13}
\frac{\partial x_\eta }{\partial E}\Delta
E_{loss} \sqrt {\frac{\Delta E_b }{\Delta E_{loss} }n_c } =
 = \Delta x_{\eta ,0} \sqrt {\frac{\Delta E_{loss} }{\Delta E_b }n_c } ,
\end{equation}
\vskip -2mm
\noindent
where $\Delta x_{\eta ,0} $ is the initial spread of closed orbits of the
beam.

In the smooth approximation the relative phase shifts of particles in their
URWs radiated in the pick-up undulator and displaced to the entrance of kick
undulators depend on theirs energy and amplitude of betatron oscillations.
If we assume that the longitudinal shifts of URWs $\Delta l < \lambda _{UR}
/ 2$, then the amplitudes of betatron oscillations, transverse horizontal
emittance of the beam and the energy spread of the beam, in the smooth
approximation, must not exceed the values

\vskip -5mm
\begin{equation}
\label{eq14}
a < < \frac{\sqrt {\lambda _{UR} \lambda _{bet} } }{\pi },
\quad
\varepsilon _x < 2\lambda _{UR} ,
\quad
\frac{\Delta \gamma }{\gamma } < \frac{\beta ^2}{\eta _c }\frac{\lambda
_{UR} }{\lambda _{bet} },
\end{equation}
\vskip -2mm
\noindent
where $\eta _c = \alpha _c - \gamma ^{ - 2}$ and $\alpha _c $ are local slip
and momentum compaction factors between undulators. Strong limitations (\ref{eq14})
to the energy spread can be overcame if, according to the decrease of the
high energy edge of the being cooled beam, a change in time of optical paths
of URWs is produced. Special elements in storage ring lattices (short
inverted dipoles, quadrupole lenses et al.) to decrease the slip [7] can be
used as well. With cooling of fraction of the beam at a time only, the
lengthening problem diminishes also as the $\Delta E / E$ now stands for the
energy spread in the part of the beam which is under cooling at the moment.

The power of the amplifier is equal to the power of the amplified URWs

\vskip -5mm
\begin {equation}
\label{eq15}
P_{ampl} = \varepsilon _{sample} \cdot f \cdot N_p / N_{kick} ,
\end{equation}
\vskip -2mm
\noindent
where $\varepsilon _{sample} = \hbar \omega _{1,\max } N_{ph} \alpha _{ampl}
$ is the energy in a sample; $N_p $, the number of particles in the ring.

The transverse selectivity of radiation (movable screen) can be arranged
with help of electro-optical elements. These elements contain crystals,
which change its refraction index while external voltage applied. This
technique is well known in optics [8]. In simplest case the sequence of
electro-optical deflector and a diaphragm followed by optical lenses, allow
controllable selection of radiation generated by different parts of the
beam.

\section{EXAMPLE}

Now let us consider an example of enhanced cooling of fully stripped
$_{207}^{82} Pb$ ion beam in the CERN LHC. The relevant parameters of the
LHC: circumference C=27 km, $f = 1.1 \cdot 10^4$, $\alpha _c = 3 \cdot 10^{
- 4}$, $\gamma = 10^3$, $M_p c^2\gamma = 192$ TeV, $\Delta \gamma / \gamma =
10^{ - 4}$, $N_p = 3 \cdot 10^9$. 1 pick-up, 10 kick undulators with
parameters $\sqrt {\overline {B^2} } = 10^5$ Gs, $\lambda _u = 100$ cm,
$M=$30 are used. The amplifier gain goes to be $\alpha _{ampl} = 10^6$.

In this case: $N_{ph} = 422$, $i_c = 0.26$mA ($N_c = i_c / ef =  \quad 1.7 \cdot
10^9)$, $\lambda _{UR,1} = 5 \cdot 10^{ - 5}$ cm, $K$=0.37, $\sigma _w = 7.7
\cdot 10^{ - 2}$cm, $E_w \cong 0.356$ V/cm, $P_{loss} = 4.35 \cdot 10^8$
eV/sec, $\Delta E_{loss} =  \quad 3.95 \cdot 10^4$ eV/rev, $\tau _u = 44.1$sec,
$P_{ampl} = 450$ W, the bandwidth of the URW $\Delta \omega / \omega \cong 1
/ K = 1 / 30 < < \Delta \gamma / \gamma $ (particles overlapped with the
URWs interacting with them), $a < 5$ mm.

\section{CONCLUSION}

We considered EOC of particle beams in storage rings for unbunched beam. EOC
in a RF bucket, peculiarities of cooling at the regime $i > i _c$, the
influence of a noise of an optical amplifier, other examples of ion
cooling in RHIC, HERA et al.  and EOC of other particles will be
considered in separate publications.

Supported by Russian Foundation of Basic Research under Grant No
05-02-17162 and by NSF.

\begin{center}
\textbf{REFERENCES}
\end{center}

[1] E.G.Bessonov, physics/0404142.

[2] A.A.Mikhailichenko and M.S. Zolotorev, Physical Review Letters, v. 71,
p.4146, 1993.

[3] A.Mikhailichenko, a Talk presented at LASER's 97, Dec 1997, Proceedings,
STS Press, McLean, VA, Ed. J.Carrol{\&}T.Goldman, p.890, ISSN 0190-4132,
1998. Also CLNS 98/1539, available at

\underline {http://ccdb3fs.kek.jp/cgi-bin/img/allpdf?199802167}.

[4] E.G.Bessonov, Proc. 18th Int. Conf. on High Energy Accelerators, HEACC
2001, March 26-30, 2001, Tsukuba, Japan,

\underline {http://conference.kek.jp/heacc2001/. Proceedings} HP. html
(P2new11); physics/0203036.

[5] E.G.Bessonov, E.V.Tkalya, Proc. First HERA-III Workshop: "The new
Frontier in Precision Lepton-Nucleon Physics", Max-Plank-Inst., Munich,
18-20 Dec. 2002; Physics/0212100; available at http:// \underline
{wwwherab.mppmu.mpg.de/hera3/Presentations.html}.

[6] E.G.Bessonov, A.A.Mikhailichenko, Preprint CLNS 01/1745, June 5, Cornell
2001, available at

\underline {http://www.lns.cornell.edu/public/CLNS/2001/}.

[7] A.Amiry, C.Pellegrini, WS on 4$^{th}$ Generation Light

Sources, SSRL 92/02, p.195.

[8] [3.3] V.J. Fowler, J. Schlafer, Applied Optics$,$ Vol.5, N10, 1657(1966).

\end{document}